\documentclass[]{aa} 
\usepackage{txfonts,epsfig,graphicx,natbib,url,twoopt,amsmath,algorithm2e}
\usepackage{color}

\newcommand{\thetabold}{\mbox{\boldmath$\theta$}}
\newcommand{\thetaboldsmall}{\mbox{\larger[-2]{\boldmath$\theta$}}}

\newcommand{\phibold}{\mbox{\boldmath$\phi$}}
\DeclareMathOperator*{\argmin}{arg\,min}
\DeclareMathOperator*{\argmax}{arg\,max}

 \usepackage[breaklinks=true,colorlinks=true,urlcolor=blue,citecolor=blue,linkcolor=blue]{hyperref} 

\bibpunct{(}{)}{;}{a}{}{,}             
\makeatletter
  \newcommandtwoopt{\citeads}[3][][]{\href{http://adsabs.harvard.edu/abs/#3}%
    {\def\hyper@linkstart##1##2{}%
     \let\hyper@linkend\@empty\citealp[#1][#2]{#3}}}
  \newcommandtwoopt{\citepads}[3][][]{\href{http://adsabs.harvard.edu/abs/#3}%
    {\def\hyper@linkstart##1##2{}%
     \let\hyper@linkend\@empty\citep[#1][#2]{#3}}}
  \newcommandtwoopt{\citetads}[3][][]{\href{http://adsabs.harvard.edu/abs/#3}%
    {\def\hyper@linkstart##1##2{}%
     \let\hyper@linkend\@empty\citet[#1][#2]{#3}}}
  \newcommandtwoopt{\citeyearads}[3][][]%
    {\href{http://adsabs.harvard.edu/abs/#3}
    {\def\hyper@linkstart##1##2{}%
     \let\hyper@linkend\@empty\citeyear[#1][#2]{#3}}}
\makeatother

\begin{document}

\title{Inference of the chromospheric magnetic field orientation in the \ion{Ca}{ii}~8542 \AA\ line fibrils}

\author{A. Asensio Ramos\inst{1,2}, J. de la Cruz Rodr\'{\i}guez\inst{3}, M. J. Mart\'{\i}nez Gonz\'alez\inst{1,2}, H. Socas-Navarro\inst{1,2}}

\institute{
 Instituto de Astrof\'{\i}sica de Canarias, 38205, La Laguna, Tenerife, Spain; \email{aasensio@iac.es}
\and
Departamento de Astrof\'{\i}sica, Universidad de La Laguna, E-38205 La Laguna, Tenerife, Spain
\and
Institute for Solar Physics, Dept. of Astronomy, Stockholm University, Albanova University Center, 10691 Stockholm, Sweden
}
             
  \date{Received ---; accepted ---} 

  \abstract
  {Solar chromospheric fibrils, as observed in the core of strong chromospheric spectral lines, extend from photospheric 
  field concentrations suggesting that they trace magnetic field lines. These images have been historically used as proxies 
  of magnetic fields for many purposes.}
  {Use statistical analysis to test whether the association between fibrils and magnetic field lines is justified.}
  {We use a Bayesian hierarchical model to analyze several tens of
    thousands of pixels in spectro-polarimetric chromospheric images of penumbrae and chromospheric fibrils.
    We compare the alignment between the field azimuth inferred from the linear polarization signals through the transverse Zeeman
    effect and the direction of the fibrils in the image.}
  {We conclude that, in the analyzed fields of view, fibrils are often well aligned with the magnetic field azimuth. 
  Despite this alignment, the analysis also shows that there is a non-negligible dispersion. In penumbral filaments,
  we find a dispersion with a standard deviation of $\sim 16^\circ$, while this dispersion goes up to
  $\sim 34^\circ$ in less magnetized regions.}
  {}

   \authorrunning{Asensio Ramos et al.}
   \titlerunning{Alignment of chromospheric fibrils and magnetic fields}
   \maketitle
%

\section{Introduction} 
The solar chromosphere can be observed in the core of strong lines
with sufficient opacity to be sensitive to the physical conditions
above the photosphere (e.g., H$\alpha$, \ion{Ca}{ii}~H \& K,
\ion{Ca}{ii}~infrared triplet, \ion{Mg}{ii}~h \& k). It can also be
observed in the \ion{He}{i} D$_3$ and $\lambda 10830$ lines which are
influenced by ultraviolet photons from the corona
(\citeads{2008ApJ...677..742C}). In active regions, these images at the
core of strong lines show a conspicuous filamentary structure
fanning out from photospheric magnetic field concentrations, suggesting
that they trace the magnetic field lines. This makes fibrils a natural proxy for the
magnetic field orientation. Similar fibrils are
observed in the umbra, penumbra and superpenumbra of sunspots with a very dynamic
behaviour (see \citeads{2013ApJ...776...56R} and
\citeads{2014ApJ...787...58Y}).

The relation between fibrils and magnetic field lines is
appealing and intuitive. However, to our knowledge, very few studies
have tried to establish this assumption. The main reason is the low polarimetric noise of the
spectropolarimetric observations required to 
obtain \emph{quantitative} information of the magnetic field vector. This
explains why the very few observational attempts to test this
assumption have been performed in penumbrae and superpenumbrae of sunspots, where
polarized signals are generally stronger. 
To this aim, \citetads{2011A&A...527L...8D}
used datasets in the \ion{Ca}{ii}~$\lambda 8542$ line whereas
\citetads{2013ApJ...768..111S} and \citetads{2015SoPh..290.1607S} used
\ion{He}{i}~$\lambda 10830$ observations. The magnetic field vector
inferred from these observations confirm, in most cases, the alignment
between fibrils and the magnetic field, although
\citetads{2011A&A...527L...8D} found cases of considerable
misalignment in their \ion{Ca}{ii} data.

In particular, \citetads{2011A&A...527L...8D} compared the visual
orientation of fibrils with the inferred azimuth obtained from Zeeman-induced linear
polarization measurements in the spectral line. The inherently noisy
Stokes $Q$ and $U$ profiles (linear polarization appears at second
order in the magnetic field in the Zeeman effect) forced them to average the Stokes
parameters along a non-negligible length of the fibrils to improve the
signal-to-noise ratio.

\citetads{2015ApJ...802..136L} studied the properties of fibrils using
a 3D numerical simulation. The authors computed magnetic field lines
starting at many seed points in the photosphere, and compared the
orientation of the field with the 3D orientation of the fibrils where
they found optical depth unity in the core of the H$\alpha$ line.
They concluded that fibrils in the simulation are mostly aligned with
the horizontal component of the magnetic field, but not necessarily
always aligned with the vertical component. More recently, 
\citetads{sykora16}, using radiative magneto-hydrodynamic simulations with a generalized Ohm’s law, have shown that the magnetic
field is often not precisely aligned with chromospheric fibrils
in places were the ambipolar diffusion is large. This is a consequence
of the slip between the field lines and the neutral species produced by
the decoupling between neutrals and ions.

\begin{figure*}
\includegraphics[width=\textwidth]{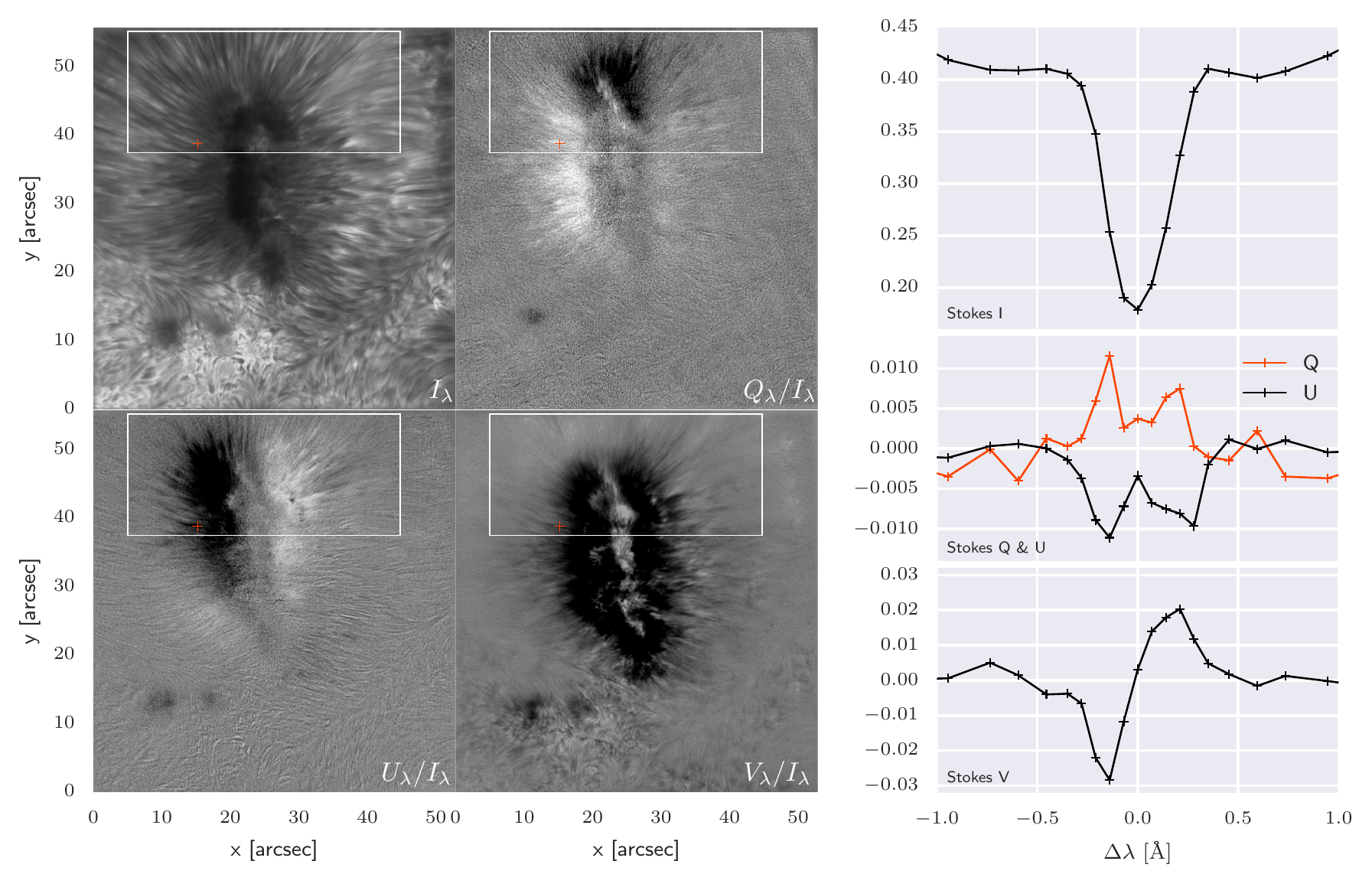}  
\caption{\emph{Left:} Clockwise the panels show Stokes $I$, $Q$, $U$
  and $V$ images at $\Delta \lambda = -140$~m\AA\ from line center in
  the \ion{Ca}{ii}~8542 line for the observation of the penumbra. The analysis has been carried out in the
  subfield indicated with a white rectangle. \emph{Right:} Full-Stokes
  spectra corresponding to the pixel indicated with a red marker in
  the FOV.}
\label{fig:obs}
\end{figure*}

From an observational point of view, this work improves over that of \citetads{2011A&A...527L...8D} by
utilizing more advanced techniques and better observations that allow us to analyze all
relevant pixels in the image, increasing the statistics to several
tens of thousands pixels, thus avoiding any averaging. The first
improvement is the application of the rolling Hough transform
(\citeads{2014ApJ...789...82C}) to estimate the direction of the
fibrils in all the pixels. The second improvement is the application
of a fully Bayesian hierarchical model \citep{gelmanHierarchical07}
for the estimation of the statistical properties of the misalignment
distribution.

\begin{figure*}
\centering
\includegraphics[width=\textwidth]{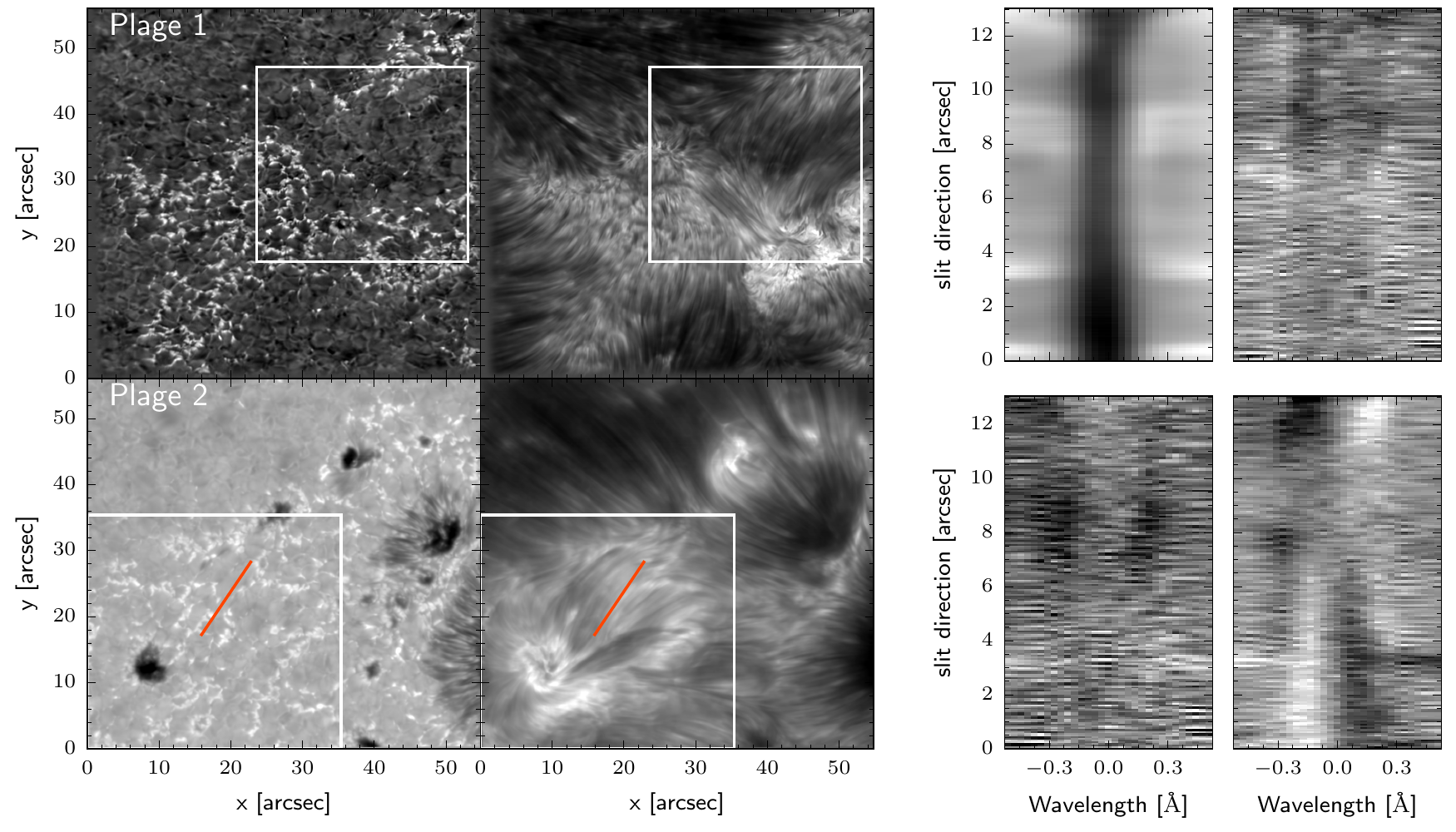}
\caption{Plage observations, marking the analyzed region with a rectangle.
The left panel shows the images in the continuum (left column) and in the core 
of the Ca \textsc{ii} 8582 \AA\ line. The right panel displays an artificial slit
along the orange line for Plage 2.}
\label{fig:plageImages}
\end{figure*}

\section{Polarimetric data}
We consider spectropolarimetric observations of a penumbra and two plages around 
AR11793 recorded on July 19 and 22, 2013. The observations on
July 19 started at 8:15 UT (plage) and 13:33 UT (penumbrae) at heliocentric distance $\mu=0.91$, 
while the observation at July 22 (plage 2) started at 08:33 UT at heliocentric distance 
$\mu=0.90$. Both datasets were obtained with the CRisp Imaging
Spectro-Polarimeter (CRISP, \citeads{2008ApJ...689L..69S}), a dual
Fabry-Perot interferometer mounted in telecentric configuration at the
Swedish 1-m Solar Telescope (SST, \citeads{2003SPIE.4853..341S}). The spatial
sampling is 0.059 arcsec/pixel. The
data were processed using the CRISPRED package
(\citeads{2015A&A...573A..40D}). The seeing on July 19 was very
good and no extra compensation for the atmospheric aberrations is 
applied. On the contrary, the Multi-Object, Multi-Frame Blind-Deconvolution
technique (MOMFBD; \citeads{1994A&AS..107..243L};
\citeads{2005SoPh..228..191V}) is applied to the data of 
July 22. The MOMFBD technique can enhance the noise even though a spatial filtering is applied to
reduce this effect. For this reason, we consider that it is advisable to use
non-corrected data for a quantitative analysis of the Stokes profiles if they
are expected to be very close or even below the noise level.
The polarimetric calibration was
performed independently for each pixel of the field of view (FOV) as
proposed by \citetads{2008A&A...489..429V}. Finally, we selected observations
very close to disk center for an easier identification of linear
polarization signals with the real azimuth of the magnetic field (thus avoiding
line-of-sight effects).

The \ion{Ca}{ii}~$\lambda 8542$ \AA\ was sampled in the range $\pm
1.75$ \AA\ from the core of the line in 21 steps of different size:
70~m\AA\ close to the Doppler core of the line, 100~m\AA\ in the wings
close to the core of the line, and 140~m\AA\ in the far wings. The
sampling is thus almost a factor 2.3 better than that of the
CRISP dataset used by \citetads{2011A&A...527L...8D} and the exposure
time is also twice as long (400 ms total integration time taking into account
the 4 modulation states). The resulting noise level is in the range $4-6\times 10^{-3}$ in
units of the continuum intensity for Stokes $Q$, $U$ and $V$.
The upper left panel of
Fig.~\ref{fig:obs} shows an example of the monochromatic image in the core of the
line for the penumbra observation, where we clearly distinguish the fibrils that are assumed to be associated
with magnetic field lines. In this study we have selected the
limb-side of the sunspot (indicated with a rectangle) because the
Stokes~$Q$ and~$U$ signals are stronger there than in any other part of
the FOV. The plage observations are summarized in Fig. \ref{fig:plageImages}. Because
the signals are lower, in this case we show an artificial slit along the orange line to
help the reader distinguish the presence of polarimetric signals. One can
easily see Stokes $V$ signals close to the footpoints, and linear polarization
in between, something that is to be expected. This figure also shows that above the
bright points the Stokes $V$ signal appears in the photospheric wings (position 3" along
the slit), but in general the polarization signals appear close to the chromospheric core of the
line.

To interpret the polarimetric signals, we consider that the magnetic
field strength is sufficiently weak that the Stokes parameters are
formed in the so-called weak-field regime
(\citeads{1973SoPh...31..299L}). In this regime, the Zeeman splitting,
$\Delta \lambda_B$, is much smaller than the Doppler broadening,
$\Delta \lambda_D$ (e.g., \citeads{2004ASSL..307.....L}). For a very
broad line such as the infrared \ion{Ca}{ii} 8542~\AA\ line, this
assumption is usually appropriate (\citeads{2013A&A...556A.115D}),
especially given that we are sensing the magnetic field in the
chromosphere, which is expected to be weaker than in the
photosphere. In this case, the weak-field approximation allows one to
compute the magnetic field vector much faster than with a
depth-stratified non-LTE inversion where even the isotopic splitting
of the \ion{Ca}{ii}~$8542$ line must be taken into account
(\citeads{2000ApJ...530..977S};
\citeads{2014ApJ...784L..17L}). Although based on quite strong
simplifying assumptions (\citeads{2004ASSL..307.....L}) a reasonably accurate
magnetic field azimuth is still obtained. Under this approximation, the
linear polarization profiles $Q$ and $U$ are given by the following
expressions (where all the quantities are constant along the
line of sight):
\begin{align}
Q_w(\lambda) &= \beta_w B_\perp^2 \left[ \frac{\partial I(\lambda)}{\partial \lambda} \frac{1}{\lambda-\lambda_0} \right]_w \cos 2 \phi, \nonumber \\ 
U_w(\lambda) &= \beta_w B_\perp^2 \left[ \frac{\partial I(\lambda)}{\partial \lambda} \frac{1}{\lambda-\lambda_0} \right]_w \sin 2 \phi,
\label{eq:QUwing}
\end{align}
which are only valid on the wings of the line, as indicated with the
subindex $w$. We choose the ranges $[-350,-140]$ m\AA\ and $[140,350]$
m\AA\ to compute these quantities, which are sufficiently far away
from the line core but some Stokes $Q$ and $U$ signal is still
measurable. Additionally, and according to \cite{quintero_ca16}, these
wing signals do still have a strong chromospheric contribution, with the response
functions peaking very high in the atmosphere.

\begin{figure*}
\includegraphics[width=\textwidth]{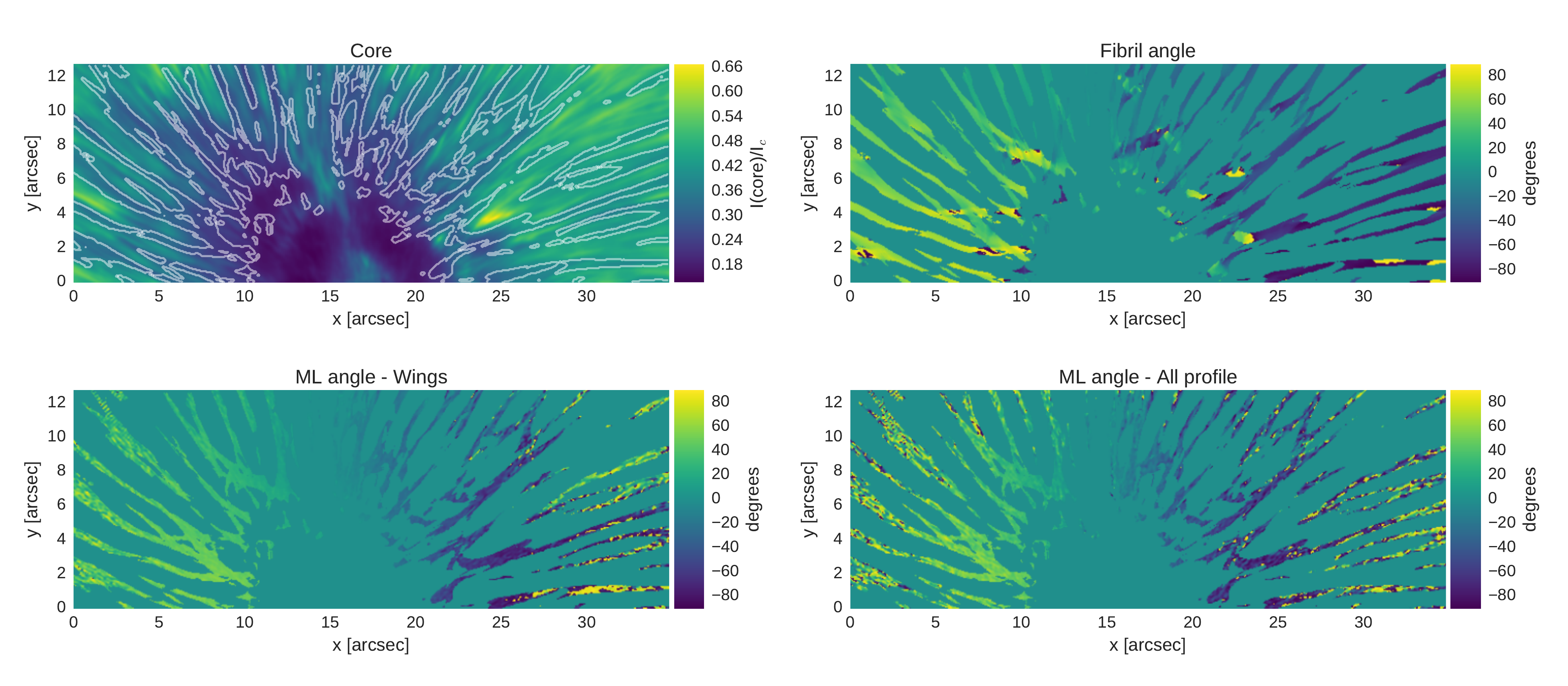}  
\caption{Upper left panel: image at the core of the Ca \textsc{ii}
  line in units of the average continuum intensity in the quietest
  region of the map. Upper right panel: estimated fibril angle using
  the RHT in the considered pixels. Lower panels: maximum-likelihood
  estimation of the azimuth angle obtained from the Stokes $Q$ and
  Stokes $U$ signal using only wavelengths on the wings (left panel)
  or the full line profile (right panel).}
\label{fig:fibrils}
\end{figure*}

In the previous equations, $\beta_w=1.63 \times 10^{-25}
\bar{G} \lambda_0^4$ is a constant that depends on the specific
spectral line of interest, with $\bar{G}$ the 
second order effective Land\'e factor for linear polarization
(cf. \citeads{2004ASSL..307.....L}) and $\lambda_0$ the central
wavelength of the line. Additionally, $B_\perp$ is the
component of the magnetic field transverse to the line-of-sight (LOS),
$\phi$ is the azimuth of the field in the plane perpendicular to the
LOS, $\lambda$ is the wavelength and $I(\lambda)$ is the Stokes $I$
profile of the spectral line. Note that the numerical factor is valid if the field is measured
in G and the wavelength in \AA. The observed Stokes $Q$ and $U$ have
been conveniently rotated so that the axis for which Stokes $Q>0$
(defined by the projection of the axis of the polarimetric analyzer on
the plane-of-the-sky) lies along the vertical direction in the maps of
Fig. \ref{fig:fibrils}.

One might also work with similar expressions that are valid on the entire line profile under more restrictive assumptions (see p. 407 of \citeads{2004ASSL..307.....L}). Such expressions are proportional to the second derivative of the intensity profile with respect to wavelength:
\begin{align}
Q(\lambda) &= -\beta B_\perp^2 \left[ \frac{\partial^2 I(\lambda)}{\partial \lambda^2}  \right] \cos 2 \phi, \nonumber \\ 
U(\lambda) &= -\beta B_\perp^2 \left[ \frac{\partial^2 I(\lambda)}{\partial \lambda^2}  \right] \sin 2 \phi,
\end{align}
where now $\beta=5.45 \times 10^{-26} \bar{G} \lambda_0^4$. All the subsequent calculations have been
obtained using Eqs.~(\ref{eq:QUwing}), which gives less noisy results, as demonstrated in
Fig. \ref{fig:fibrils}.

Using the formulation worked out by \citetads{2012MNRAS.419..153M}, we
estimated the maximum-likelihood value of the azimuth at each
considered pixel.  The results are displayed in the lower panels of
Fig.~\ref{fig:fibrils}. These results suggest an overall good
alignment between azimuth angle and fibril direction but no
information on the observational uncertainties has been considered in
the analysis.

\section{Detection of fibrils}
To compare the field orientation inferred from the polarimetry to that
from the fibrils, we need to have a reliable estimate of the fibril
direction in all relevant pixels. To this end, we make use of the
recently presented rolling Hough transform (RHT,
\citeads{2014ApJ...789...82C}), developed for detecting fibrils and
estimating their direction in images of the interstellar
medium\footnote{We use the Python code publicly available at
  \texttt{https://github.com/seclark/RHT}.}.  The RHT is a
generalization of the standard Hough transform and is obtained after
the following steps:
\begin{itemize}
\item A smoother version of the image is computed and then substracted
  from the original image. The smoothing is obtained using a top-hat
  kernel of a certain width $D_K$, which acts as a high-pass filter to
  suppress large scales.
\item The resulting image is then thresholded and binarized to create
  a bitmask.
\item A disk of a certain diameter $D_W$ is extracted at every point
  in the image and the standard Hough transform is computed in each
  disk.
\item Finally, the Hough transform for each disk is thresholded at a
  certain level $Z$ to make sure that only obvious fibrils are
  detected as such.
\end{itemize}
As a consequence of its definition, the RHT is specially indicated to
detect structures whose length is equal or longer than $D_W$ and with
a brightness contrast larger than the threshold $Z$. This is specially suitable
for the relatively diffuse images of the core of the Ca \textsc{ii}
line (see \citeads{2014ApJ...789...82C} for more details).

The output of the rolling Hough transform is the discretized function
$R(\theta,x,y)$, which is defined at each pixel position $(x,y)$ and
angle $\theta$. This function describes, for each disk of diameter
$D_W$ centered at position $(x,y)$, the angles of the dominant linear
structures. A visualization of the linear structures can be obtained
by computing the backprojection $R(x,y)$, defined as:
\begin{equation}
R(x,y) = \int R(\theta,x,y) \, \mathrm{d}\theta.
\end{equation}

Assuming that there is a preferential linear structure, the dominant
azimuth at each pixel position is obtained by computing the circular
statistics average at each pixel:
\begin{equation}
\phi_\mathrm{RHT} = \frac{1}{2} \arctan \frac{\int R(\theta,x,y) \sin (2\theta) \mathrm{d}\theta}{\int R(\theta,x,y) \cos (2\theta) \mathrm{d}\theta}.
\end{equation}
The upper right panel of Fig. \ref{fig:fibrils} displays the angle of 
each linear structure in the penumbra as obtained applying the RHT to the
image on the upper left panel of the same figure. We found that using
$D_K=10$, $D_W=55$ and $Z=0.7$ gives very good results but the results are
not strongly dependent on small variations around these values. The reference
angle is along the vertical direction. The umbra is removed
from the azimuth map using a mask. In summary: we select dark structures
in the map of the core brightness, we avoid the darkest regions in the
continuum image to remove the umbra and we only choose points with a
backprojection larger than 20\% of the maximum (so that we only choose
structures with a well-defined linear appearance). Once the dark filamentary
structures are selected, the bright features are also trivially obtained as 
just the negative of the chosen mask (always discarding the umbra). We considered both the bright and
dark structures in our analysis.

\begin{figure}
\includegraphics{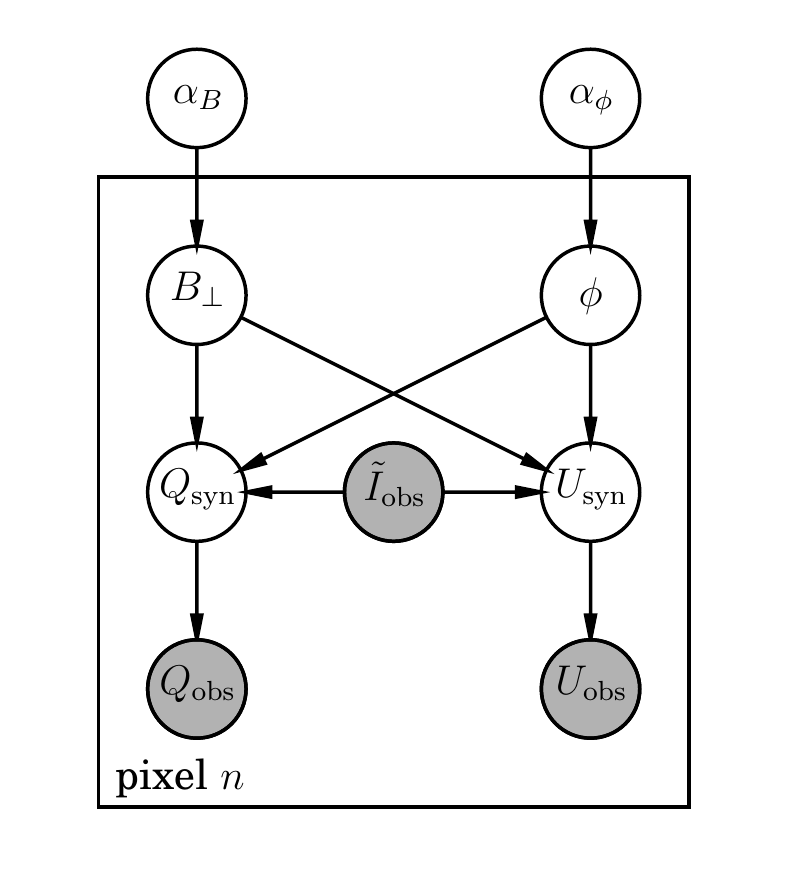}  
\caption{Graphical model representing the conditional dependences among the variables of the
statistical model.}
\label{fig:graphical_model}
\end{figure}

\section{Inference of magnetic field azimuth}
\subsection{Bayesian hierarchical model}
Under the presence of uncorrelated Gaussian noise, it is
straightforward to write down the generative model used for explaining
the linear polarization in the wings of the Ca \textsc{ii} line at the
$j=1,\ldots,N_\lambda$ sampled wavelengths and for a given pixel
$i=1,\ldots,N_\mathrm{pix}$. We modify Eqs.~(\ref{eq:QUwing}) to
include the fibril angle estimated with the RHT and rechristen the
azimuth to represent the misalignment between the magnetic field
azimuth and the fibril direction:
\begin{align}
Q_{ij} &= \beta_w {B_\perp}_i^2 \tilde{I}_{ij} \cos \left[ 2 (\phi_i + \phi_{\mathrm{RHT},i}) \right] + \epsilon_{Q,ij}, \nonumber \\ 
U_{ij} &= \beta_w {B_\perp}_i^2 \tilde{I}_{ij} \sin \left[ 2 (\phi_i + \phi_{\mathrm{RHT},i}) \right] + \epsilon_{U,ij},
\label{eq:generative_model}
\end{align}
where, for the sake of a simpler notation, we use
$\tilde{I}(\lambda)=(\partial I/\partial
\lambda)/(\lambda-\lambda_0)$, with the numerical derivatives being calculated
using a quadratic Lagrangian interpolation. Note that, when $\phi_i=0$, one can safely state that the
magnetic field is along the fibril direction.

We acknowledge that our treatment of noise is somehow simplified. We assume that the noise contributions,
$\epsilon_{Q,ij}$ and $\epsilon_{U,ij}$, are Gaussian-distributed
random variables with zero mean and standard deviation $\sigma_n$. The
standard deviation is estimated from the continuum wavelengths on the
observations, where the linear polarization signal is expected to be
zero. We find $\sigma_n \sim 5-7 \times 10^{-3} I_c$, where $I_c$ is the
continuum intensity. Under this framework, we are assuming that all sources of error (photon
noise, uncertainty in the estimation of the fibril angle, fringes and any other systematic effect) are 
absorbed in these zero mean
errors with variances empirically obtained from the data. Another particularity of our
generative model is that
it depends on an observed (and, consequently, noisy) quantity, $\tilde{I}(\lambda)$. According to
\cite{asensio_manso11}, in such a case the noise variance should take this into account. For the moment, 
we neglect this second order effect in this work and we defer a more elaborate Bayesian treatment of
all systematic and random uncertainties for a future publication.

If we analyze a single pixel $i$, the information about ${B_\perp}_i$
and $\phi_i$ that can be extracted from the observations is summarized
in the posterior distribution:
\begin{align}
p(\phi_i,{B_\perp}_i|\tilde{\mathbf{I}}_i,\mathbf{Q}_i,\mathbf{U}_i) = p(\tilde{\mathbf{I}}_i,\mathbf{Q}_i,\mathbf{U}_i|\phi_i,{B_\perp}_i) p({B_\perp}_i) p(\phi_i),
\end{align}
where
$p(\tilde{\mathbf{I}}_i,\mathbf{Q}_i,\mathbf{U}_i|\phi_i,{B_\perp}_i)$
is the likelihood associated to pixel $i$ (with
$\tilde{\mathbf{I}}_i$, $\mathbf{Q}_i$ and $\mathbf{U}_i$ vectors
containing all the observed wavelengths for the pixel). Likewise,
$p({B_\perp}_i)$ and $p(\phi_i)$ are priors that are assumed to be
independent for all pixels. Applying this scheme to all pixels in the
FOV, one may estimate the diversity in the misalignments by comparing
their posteriors.

However, it is useful to make the assumption that all pixel share a
common prior and use a hierarchical model to put together all the
partial and uncertain information from each pixel to constrain this
prior (e.g., \citeads{2014MNRAS.439L..31B}). In other words, even
though each pixel is characterized by ${B_\perp}_i$ and $\phi_i$, we
put a common parametric prior that depends on the set of
hyperparameters $\alpha_\phi$ and $\alpha_B$. These priors are shared by all pixels in
the field-of-view (FOV).  Using simple rules of probability calculus,
it is easy to write the posterior distribution for all pixels
simultaneously:
\begin{align}
p(\phibold,\mathbf{B}_\perp,\alpha_\phi,\alpha_B|\tilde{\mathbf{I}},\mathbf{Q},\mathbf{U}) &= p(\tilde{\mathbf{I}},\mathbf{Q},\mathbf{U}|\phibold,\mathbf{B}_\perp) \nonumber \\
& p(\mathbf{B}_\perp|\alpha_B) p(\alpha_B) p(\phibold|\alpha_\phi) p(\alpha_\phi),
\label{eq:full_posterior}
\end{align}
which is trivially inferred from the graphical model displayed in
Fig.~\ref{fig:graphical_model}. In the previous equation,
$\tilde{\mathbf{I}}$, $\mathbf{Q}$ and $\mathbf{U}$ contain all the
observations for all pixels.  As seen from the previous expression,
the model includes a hierarchical prior for $\phibold$, which is made
dependent on the set of hyperparameters $\alpha_\phi$ (over which we
set another prior to be consistent with the Bayesian framework). Given
the assumption of uncorrelated noise in all pixels, the likelihood can
be written as (see Appendix \ref{sec:appendix}):
\begin{equation}
p(\tilde{\mathbf{I}},\mathbf{Q},\mathbf{U}|\phibold,\mathbf{B}_\perp) = \prod_{i=1}^{N_\mathrm{pix}} \mathcal{L}_{Q,i}(\phi_i,{B_\perp}_i) \, \mathcal{L}_{U,i}(\phi_i,{B_\perp}_i),
\end{equation}
with 
\begin{align}
\mathcal{L}_{Q,i}(\phi_i,{B_\perp}_i) &= \prod_{j=1}^{N_\lambda} p(\tilde{I}_{ij},Q_{ij}|\phi_i,{B_\perp}_i) \nonumber \\
\mathcal{L}_{U,i}(\phi_i,{B_\perp}_i)&= \prod_{j=1}^{N_\lambda} p(\tilde{I}_{ij},U_{ij}|\phi_i,{B_\perp}_i) \nonumber.
\end{align}


Given that our interest is to obtain statistical information about the
azimuth in the whole FOV and we are not really interested in their
specific values for individual pixels, we marginalize $\phibold$ and
$\mathbf{B}_\perp$ from the posterior distribution of
Eq. (\ref{eq:full_posterior}):
\begin{equation}
p(\alpha_\phi,\alpha_B|\tilde{\mathbf{I}},\mathbf{Q},\mathbf{U}) = 
\int \mathrm{d}\phibold \mathrm{d}\mathbf{B}_\perp p(\phibold,\mathbf{B}_\perp,\alpha_\phi,\alpha_B|\tilde{\mathbf{I}},\mathbf{Q},\mathbf{U}).
\end{equation}

\begin{figure*}
\centering
\includegraphics[width=\textwidth]{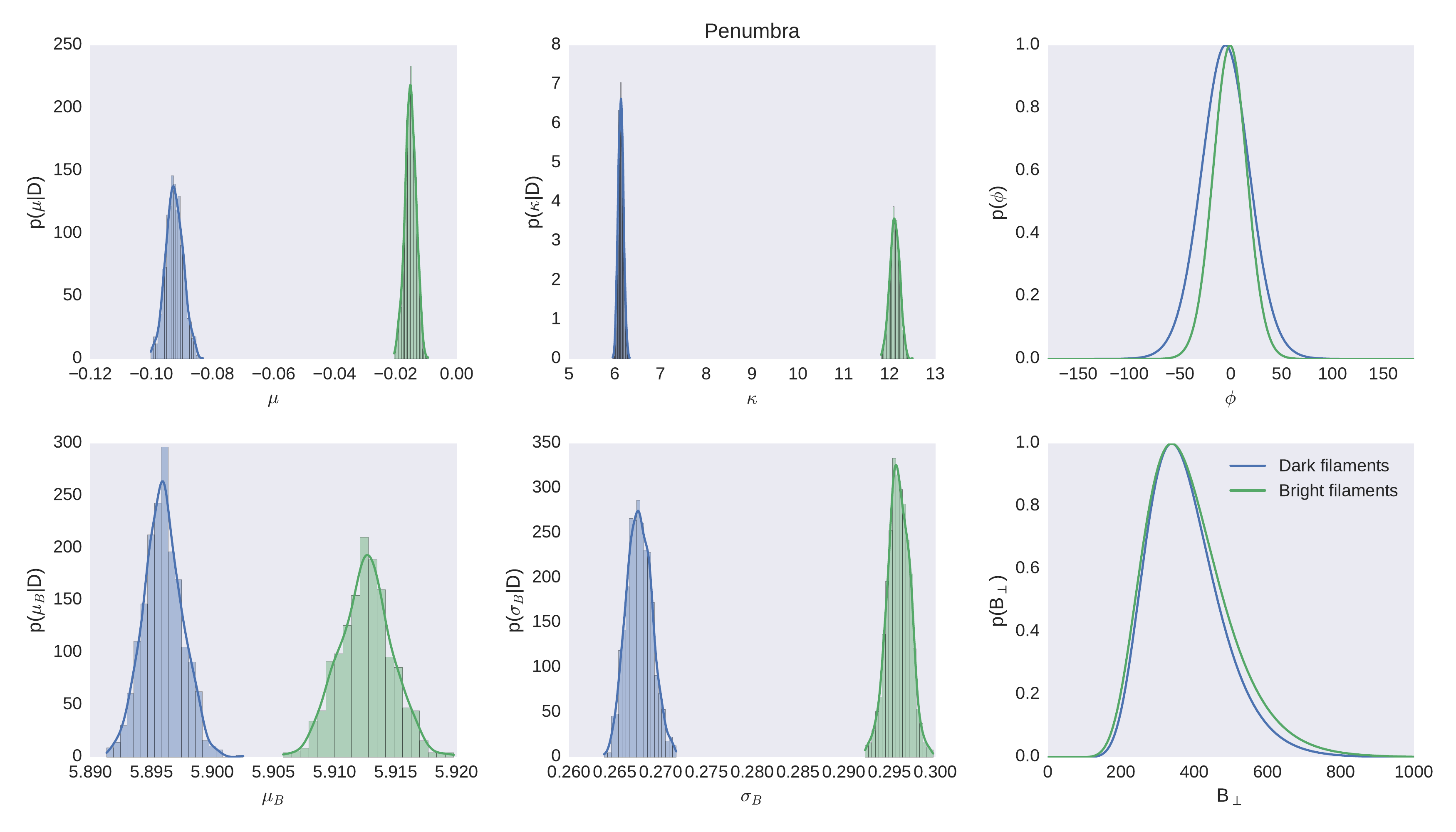}
\caption{The first and second columns show the marginal posterior for the hyperparameters of the von-Mises
  prior for the misalignment (upper panels) and the log-normal prior for the perpendicular
  component of the magnetic field (lower panels) for the case of the penumbra. The plots show a 
  histogram and a kernel density estimation using the samples. The rightmost column 
  shows the Montecarlo estimate of the prior distribution for all considered pixels for both
  parameters.}
\label{fig:hyperparameters}
\end{figure*}

\subsection{Priors and hyperpriors}
The problem is fully defined once we set the parametric priors
$p(\phi_i|\alpha_\phi)$ and $p(B_\perp|\alpha_B)$, and the hyperpriors
$p(\alpha_\phi)$ and $p(\alpha_B)$.
Concerning the prior for the misalignment, we choose a von-Mises distribution, which is naturally defined on the interval
$[-\pi,\pi]$ and is also naturally periodic. The von-Mises distribution
is one of the most used generalizations of the normal distribution for
circular statistics and is quite flexible. Its functional form is:
\begin{equation}
p_\mathrm{VM}(\phi_i|\mu,\kappa) = \frac{1}{2\pi I_0(\kappa)} \exp \left[ \kappa \cos \left( \phi_i-\mu\right)\right],
\end{equation}
which depends on the set of hyperparameters $\alpha_\phi=(\mu,\kappa)$. In
the previous expression, $I_i(\kappa)$ is the modified Bessel function
of the first kind \citepads{1972hmfw.book.....A}. As $\kappa$ increases, the 
von-Mises distribution approaches a normal distribution with the same mean and standard deviation
$\sigma=\sqrt{2}/\kappa$. To finalize, a suitable
Jeffreys'-type hyperprior $p(\mu,\kappa)$ is given by
\citep{dowe96}:
\begin{equation}
p(\mu,\kappa) \propto \left[ \kappa A(\kappa) \frac{\mathrm{d}A(k)}{\mathrm{d}k} \right]^{1/2},
\end{equation}
with $A(\kappa)=I_1(\kappa)/I_0(\kappa)$. However, we have checked that the more standard Jeffreys' prior 
$p(\mu,\kappa) \propto \sigma^{-1}$ also works equivalently in our case \citep[e.g.,][]{gregory05}.
For computational reasons, we have used the almost equivalent inverse Gamma prior
\begin{equation}
\mathrm{IG}(\sigma;\alpha,\beta) = \frac{\beta^\alpha}{\Gamma(\alpha)} \sigma^{-\alpha-1} \exp\left(\frac{-\beta}{\sigma} \right)
\end{equation}
which converges to the Jeffreys' prior when $\alpha \ll 1$ and $\beta \ll 1$.

To complete the problem, we choose a log-normal distribution for $B_{\perp,i}$:
\begin{equation}
p_\mathrm{LN}(B_{\perp,i}|\mu,\kappa) = \frac{1}{B_{\perp,i} \sqrt{2\pi} \sigma_B} \exp \left[ -\frac{\left(\log B_{\perp,i}-\mu_B \right)^2}{2\sigma_B^2}\right],
\end{equation}
which is parameterized by the location $\mu_B$ and scale $\sigma_B$ parameters, $\alpha_B=(\mu_B,\sigma_B)$.
Additionally, we set a standard Jeffreys' prior $p(\mu,\kappa) \propto \sigma_B^{-1}$ through an inverse Gamma prior. The log-normal distribution
naturally puts zero probability to $B_{\perp,i}=0$, which arises naturally from a 
non-pathological vector field in three dimensions.

\subsection{Variational method}
The computation of the marginalization integral can be potentially carried out using
Markov Chain Monte Carlo techniques. However, given the large number
of pixels we want to analyze, the integration becomes very time
consuming. For this reason, we use an automatic variational approximation,
as included in the Stan software \citep{kucukelbir_autovar_stan15}. Variational inference relies on
using a simpler parametric distribution to approximate the posterior distribution. 
In short, assume that our aim is to approximate the posterior distribution $p(\thetabold|\mathbf{X})$,
where $\thetabold$ is the vector of parameters and $\mathbf{X}$ are the observations. If one considers
the family $q(\thetabold|\phibold)$ of probability densities parameterized by the vector $\phibold$, it is possible to obtain
an approximation to the posterior by computing the value of the parameters that give a smaller value
of the Kullback-Leibler divergence, $D_\mathrm{KL}$, between the two distributions\footnote{We remind that the Kullback-Leibler divergence
is a a measure of the difference between two probability distributions $p(x)$ and $q(x)$, and it is
given by $D_\mathrm{KL}=\int \mathrm{d}x p(x) [\log p(x) - \log q(x)]$.} \citep[e.g.,][]{bishop06}:
\begin{equation}
\argmin_{\phibold} D_\mathrm{KL}[q(\thetabold|\phibold) \parallel p(\thetabold|\mathbf{X})].
\end{equation}
Since the Kullback-Leibler divergence usually lacks a closed form, it is customary to maximize a slightly different
problem, in which the evidence lower bound ($L$) appears:
\begin{equation}
\argmax_{\phibold} L(\phibold) = \mathbb{E}_{q(\thetaboldsmall)} \left[ \log p(\thetabold,\mathbf{X}) \right] -
\mathbb{E}_{q(\thetaboldsmall)} \left[ \log q(\thetabold|\phibold) \right],
\end{equation}
where $\mathbb{E}_{q(\theta)}[x]$ is the expectation value of $x$ over the distribution $q(\theta)$.
The variational approximation greatly simplifies the problem and allows it to scale
very well when the number of observations and/or variables increases. We use the 
implementation of the variational approximation included in the Stan package \citep{stan16} \footnote{The Stan Version 2.10.0 package
used in this paper can be found in \texttt{http://mc-stan.org} and the Stan code
used in this work can be found in \texttt{http://github.com/aasensio/fibrilMisalignment}.}.


\section{Results}

\begin{figure*}
\centering
\includegraphics[width=\textwidth]{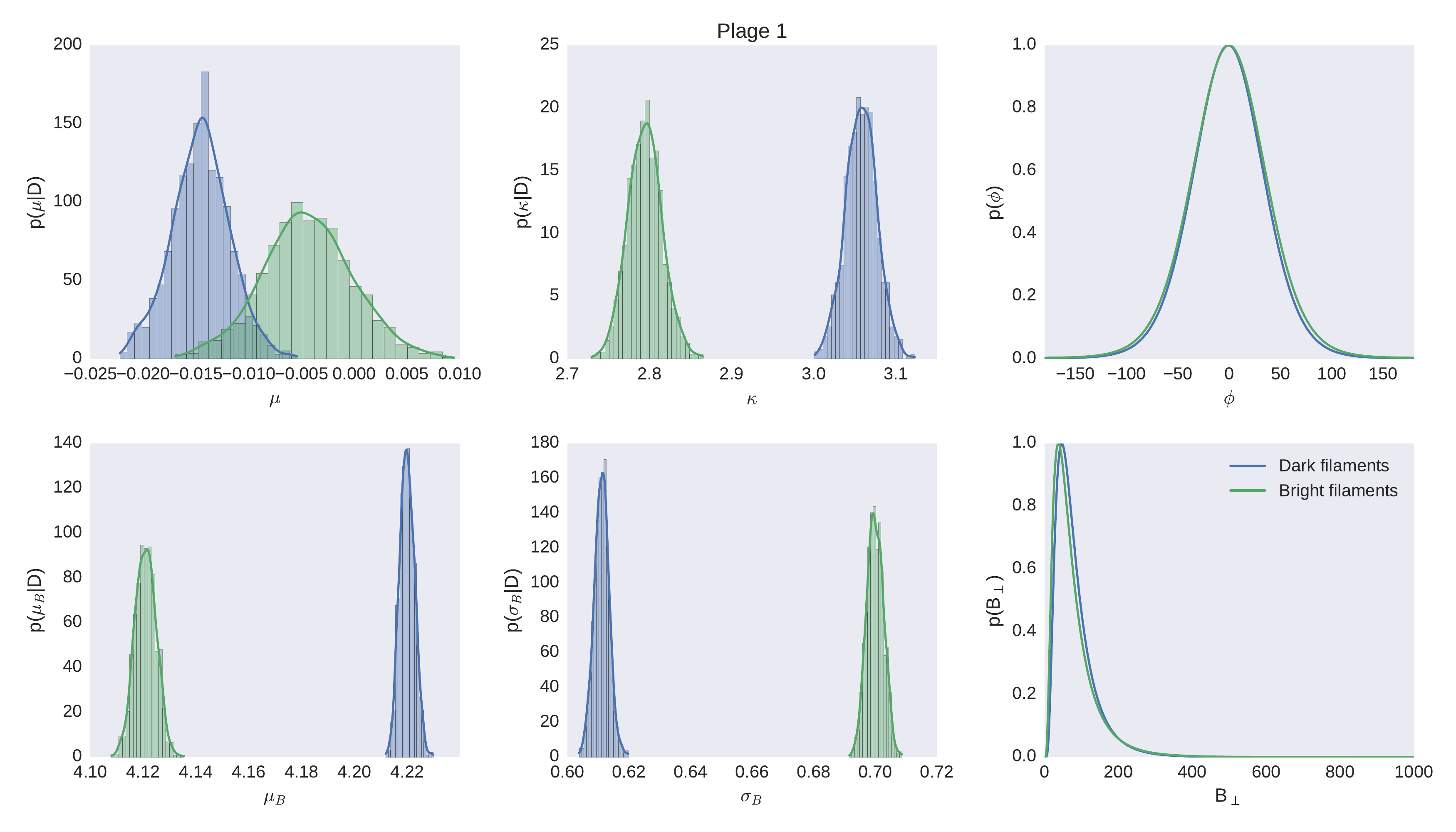}\vspace{0.5cm}
\includegraphics[width=\textwidth]{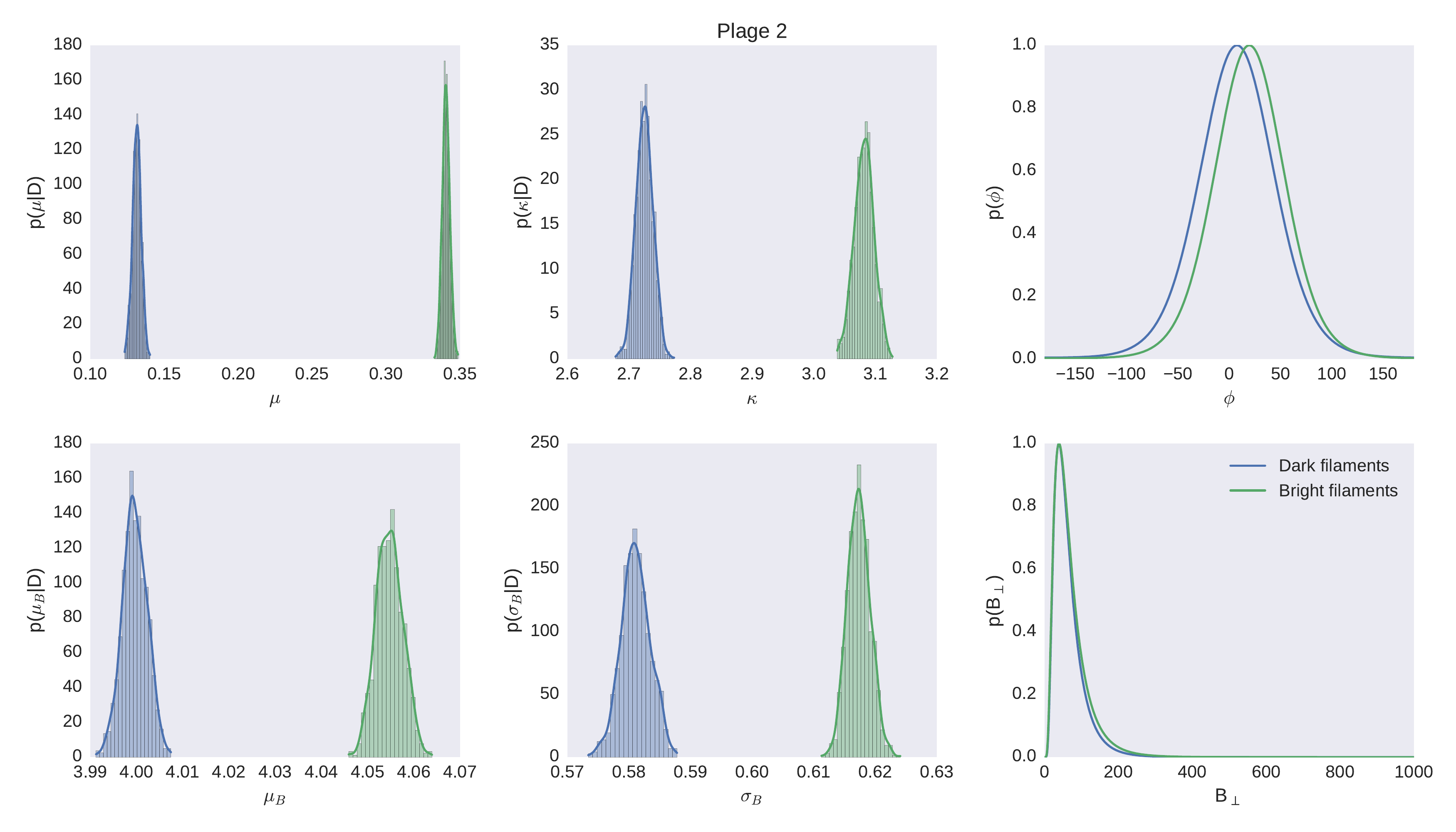}
\caption{Same as Fig. \ref{fig:hyperparameters} but for the two plages cases considered.}
\label{fig:hyperparametersPlages}
\end{figure*}

\begin{figure*}
\centering
\includegraphics[width=0.32\textwidth]{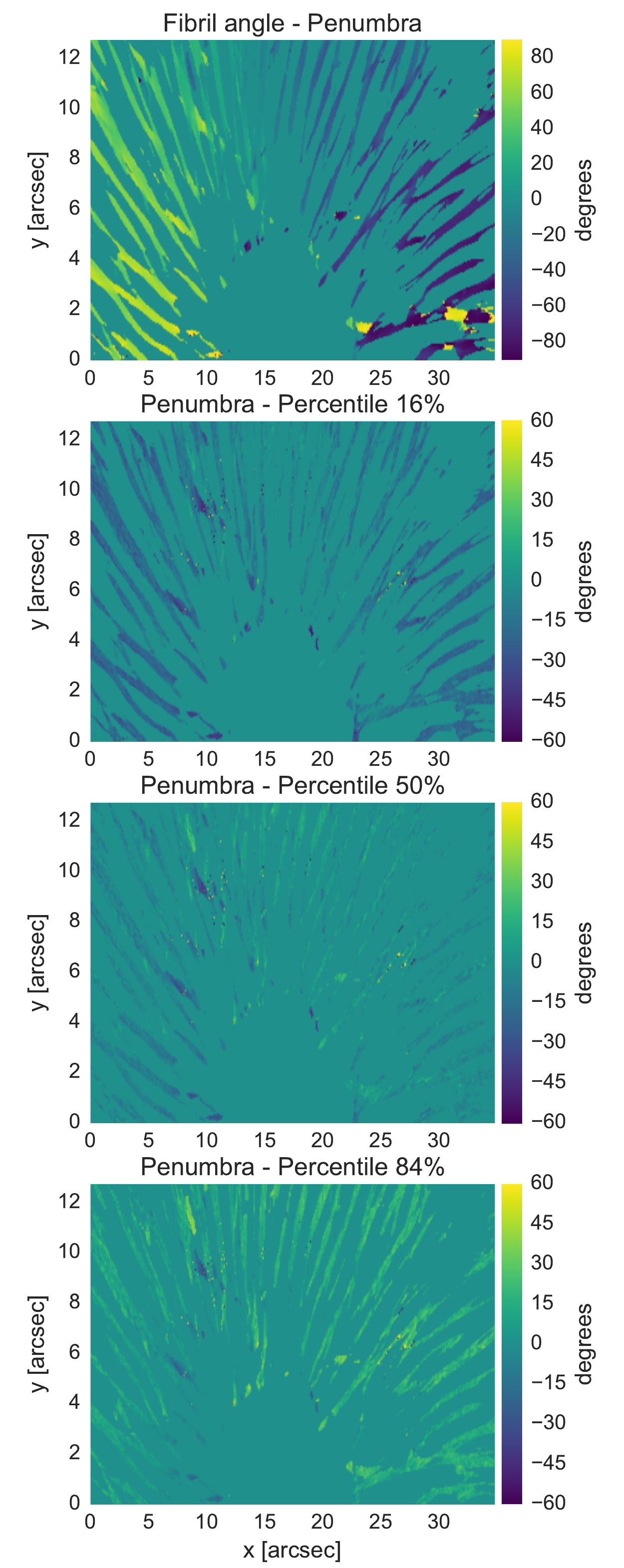}%
\includegraphics[width=0.32\textwidth]{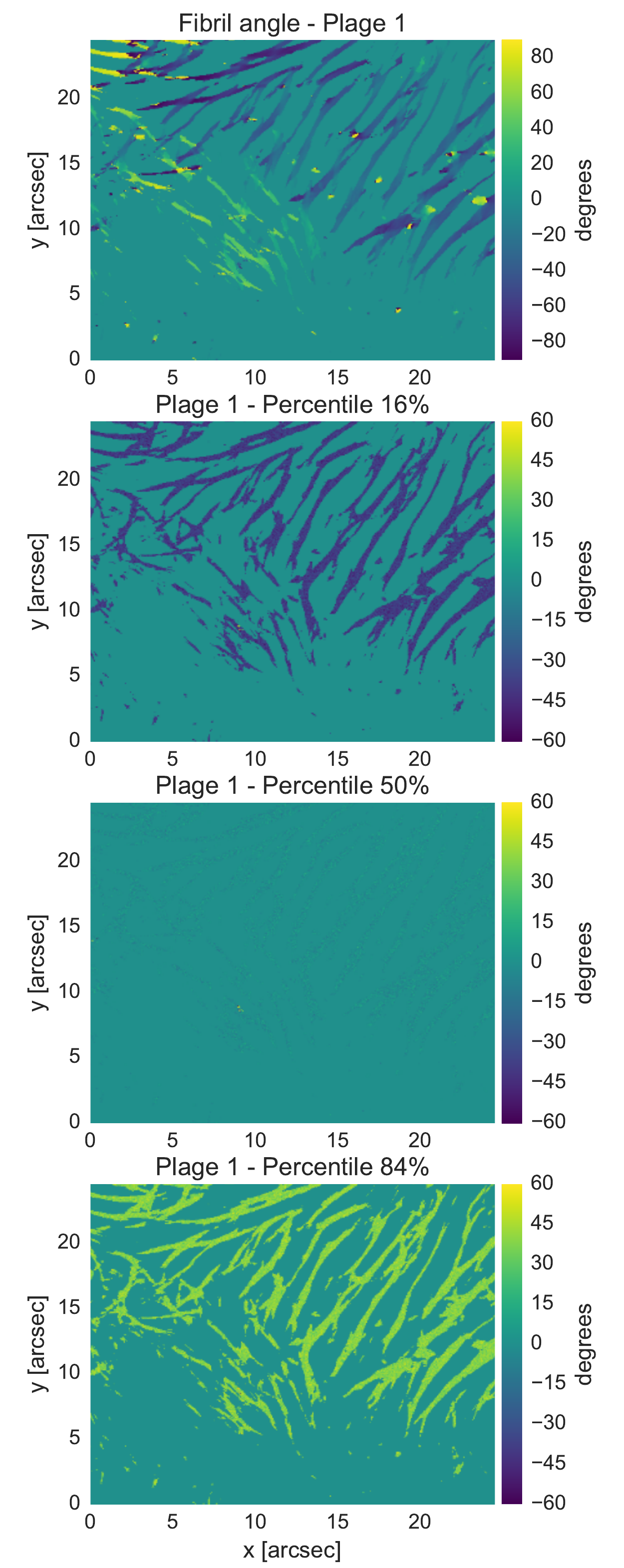}%
\includegraphics[width=0.32\textwidth]{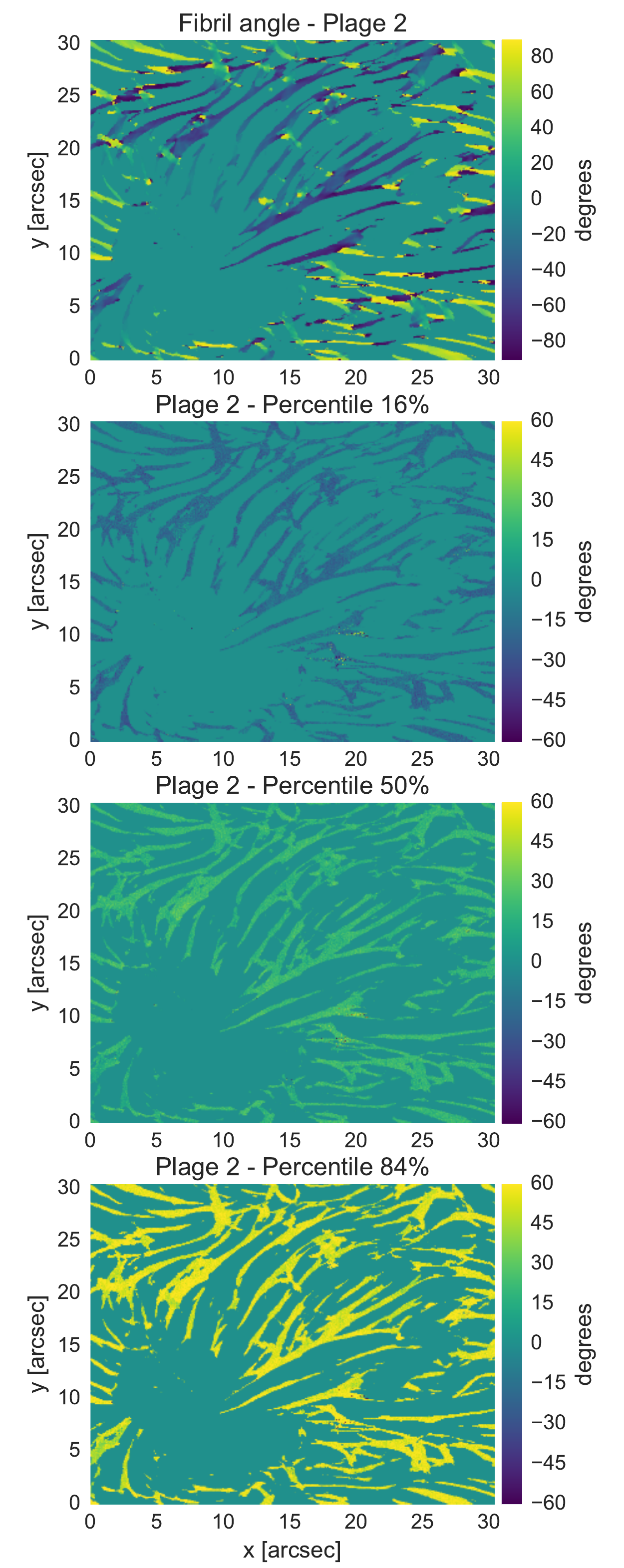}
\caption{The upper row display the fibril angle inferred from the bright structures in the images at the core of the line. The second, third
and last rows display the 16, 50 and 84 percentiles of the distribution of inferred misalignments for each
pixel.}
\label{fig:maps}
\end{figure*}

\subsection{Penumbra}
Our inference for the penumbra is based on a total of $\sim 3 \times 10^4$ pixels of the
map, both for the bright and dark structures. The marginal distributions for the
hyperparameters of the von-Mises and log-normal distributions are displayed in the first
and second columns of Fig.~\ref{fig:hyperparameters}. Note that the marginal posteriors for
the hyperparameters are very well defined for both the misalignment
and the perpendicular component of the magnetic field. The distributions point 
to a slight statistical misalignment overall
between the azimuth of the field and the fibril direction of less than
$\sim 1^\circ$ for the bright structures, and less than $\sim 6^\circ$ for the dark ones. 
Although statistically relevant, it seems rather
unimportant given that it is probably within the uncertainty in the
determination of the fibril direction. Concerning the field, it points
towards a median value for the perpendicular field of $e^{\mu_B} \sim 370$ G, compatible
in both cases. This suggests that bright and dark structures are located in regions of relatively
smooth magnetic field.

Motivated by the well-constrained values of the hyperparameters, it is advisable 
to use these samples to produce a Montecarlo estimate of the prior distribution
over the whole field of view that we used. To this end, we compute the averaged 
distribution for the misalignment and field using:
\begin{align}
\langle p(\phi) \rangle &= \frac{1}{N} \sum_{i=1}^{N} p(\phi|\alpha_{\phi,i}) \, , \nonumber \\
\langle p(B_\perp) \rangle &= \frac{1}{N} \sum_{i=1}^{N} p(B_\perp|\alpha_{B_\perp,i}) \, ,
\end{align} 
where $N$ is the number of samples obtained from the variational approximation.
The results are shown in the rightmost column of Fig.~\ref{fig:hyperparameters}. 
Given that the values of $\kappa$ are large, the results are almost Gaussian. The dark
structures show a mean of $\sim -6^\circ$ and a standard deviation of $\sim 23^\circ$, while
the bright structures display a mean of $\sim -1^\circ$ and a standard deviation of $\sim 16^\circ$. 
In other words, we find that the probability
of having a misalignment larger than 16$^\circ$ in the bright filaments is only
32\%. Additionally, the probability of finding a misalignment larger
than 32$^\circ$ goes down to 5\%. A similar reasoning is applied to the dark structures. It is important to note, however,
that these probabilities are all referred to the penumbra region observed
here.

It is true that the results displayed in Fig. \ref{fig:hyperparameters} 
depend on the number of pixel considered, with
the distribution of hyperparameters slowly converging to a certain value 
when $N_\mathrm{pix}$ increases. The convergence of the means, $\mu$ and $\mu_B$,
go roughly as $N_\mathrm{pix}^{1/2}$. This is the well-known behavior of the convergence
of the mean when adding many samples. However, the convergence of the 
hyperparameters related with the dispersion is much slower. The reason is that
these are already second-order statistics, and their uncertainty decreases as
$N_\mathrm{pix}^{1/4}$. As a consequence, combining 3$\times$10$^4$ pixels reduces
the uncertainty in the mean by a factor $\sim 170$ with respect to the single-pixel case.
Meanwhile, the uncertainty in the width of the distribution decreases only by a factor $\sim 13$.
In other words, we can safely state that the mean value of the misalignment is close to zero,
but our certainty in the dispersion is smaller. We have checked this experimentally by doing the
analysis with several number of pixels from 1 to 3$\times$10$^4$ and verifying that the
value of $\mu$ rapidly converges toward 0, while the value of $\kappa$ is probably representative
but not yet fully converged. 

Using a common prior for all pixels introduces a shrinkage effect that pushes
all inferred misalignments to share a common prior. This effect is seen in the leftmost
column of Fig. \ref{fig:maps}, where we show the percentiles 16, 50 and 84 for the misalignment
for all considered pixels, together with the map of fibril angles inferred with the RHT. 
The median value of the misaligmnent for all pixels is around zero for almost all
fibrils, except in some specific locations. Additionally, the percentile 16 shows negative
values almost all over the field of view, while the percentile 84 displays positive values.
Of interest are the regions at (10$"$,9$"$), (7$"$,6$"$) and (8$"$,1$"$), that display 
a negative misalignment in all percentiles. The misalignments can be real or they can
be produced by an incorrect estimation of the geometrical alignment of the fibril. We note
that all these cases are fibrils that are not strictly along the remaining penumbra filaments,
but are linking two parallel fibrils. 
Finally, the smooth appearance of the maps of Fig. \ref{fig:maps} suggest that the dispersion found in
the distributions of misalignments of Fig. \ref{fig:hyperparameters} is intrinsic, and
not produced by non-converged results for $\kappa$.

In other words, it is sure that the \emph{average} misalignment
is compatible with zero, but it is still unclear whether the dispersion found is produced
by noise or it is real.

\subsection{Plage}
Although checking for the alignment of chromospheric fibrils and magnetic fields in penumbrae 
is interesting, it turns out to be even more important to check for this alignment
in less magnetized regions. According to the recent simulations of \cite{sykora16},
ambipolar diffusion can often produce misalignments between the high-density
weakly ionized fibrils and the magnetic field. For this
reason, we also analyze regions above plages, whose results are displayed
in Fig. \ref{fig:hyperparametersPlages} for the two cases considered in this
work. The first plage contains $\sim$3$\times$10$^4$ pixels for the bright filaments, 
and $\sim$5$\times$10$^4$ pixels for the dark ones, while in the
second one we increased the number of points to $\sim$6$\times$10$^4$ for the bright
structures and $\sim$5$\times$10$^4$ for the dark ones. The inferred
hyperparameters indicate that the field is almost aligned with the bright and dark structures in the first case.
On the contrary, it turns out to be slightly misaligned ($\sim 19.5^\circ$ for bright structures
and $\sim 7.5^\circ$ for dark ones) in the second case. Both
share roughly the same uncertainty in the misalignment of $\sim34^\circ$. Given that
the Stokes $Q$ and $U$ signals in these regions have lower amplitudes than in the penumbra,
we cannot discard that the estimated uncertainty in the misalignment can be reduced by
adding many more pixels because it is still dominated by the presence of noise.
Concerning the magnetic field perpendicular to the line-of-sight, the results consistently
indicate that they are much smaller than in the penumbra, with median values 
equal to $e^{\mu_B} \sim 60$ G.

The shrinkage effect of the hierarchical model is demonstrated for the plage
case in the middle and right columns of Fig. \ref{fig:maps}. We find no
relevant regions in the maps with a strong misalignment.

\section{Discussion and conclusion}
To put these results in the context of previous work, let us recall that
\citetads{2011A&A...527L...8D} obtained azimuths of the field
(averaged along each fibril) that were, on average, well aligned with
the direction of the fibrils. However, they also found a non-negligible
fraction of the tens of cases analyzed where strong misalignments were
observed, in some cases close to 90$^\circ$. Our results do not
indicate the presence of such strong differences, at least
statistically. The discrepancy might be due to the presence of noise
(\citeads{2012A&A...543A..34D}), which affects the maximum likelihood
estimation of the azimuth used by \citetads{2011A&A...527L...8D}, or
to the fact that the regions are different. 
Although less likely (but possible), the findings of \citetads{2015ApJ...802..136L} using
3D simulations seem to indicate that fibrils in H$\alpha$ do not necessarily trace
the vertical component of the magnetic field, at least not where
$\tau_\mathrm{\lambda_0} = 1$. Therefore misalignments may occur in
observations close to the limb, where Stokes~$Q$ \& $U$ signals would
originate from the vertical component of the field due to projection
effects. One of the datasets used by \citetads{2011A&A...527L...8D} is
at heliocentric distance $\mu=0.41$. In the present work, the penumbra median
misalignments could also produced by \emph{elevation} effects
along the fibrils. The potential misalignments that we find in this work (although the median value is
very close to zero) might be compatible with the simulations
of \cite{sykora16}. Therefore,
observations with better signal-to-noise ratio are needed to 
observationally quantitatively pin down the importance of 
ambipolar diffusion producing strong misalignments between 
fibrils and the magnetic field. 

Our results are also in good agreement with the findings of
\citetads{2013ApJ...768..111S} who used the \ion{He}{i}~$\lambda
10830$ line to measure the alignment of fibrils in the surrounding of
a sunspot. Given that polarization in the \ion{Ca}{ii}~$\lambda 8542$
line can be modeled in active regions using (only) the Zeeman effect
(\citeads{2010ApJ...722.1416M}) and it does not suffer from extra
ambiguities in the azimuth derived from the Hanle effect and
scattering polarization, our results also reinforce the findings of
\citetads{2013ApJ...768..111S} but (in our case) using a diagnostic
from a completely different formation mechanism both for the line
(different atom, optically thick) and for the polarization (Zeeman
induced).

\appendix
\section{Likelihood}
\label{sec:appendix}
Because all observed pixels and wavelengths are assumed to be uncorrelated, we can factorize the likelihood
as follows:
\begin{align}
p(\tilde{\mathbf{I}},\mathbf{Q},\mathbf{U}|\phibold,\mathbf{B}_\perp) = \prod_{i=1}^{N_\mathrm{pix}} \prod_{j=1}^{N_\lambda} p(\tilde{I}_{ij},Q_{ij}|\phi_i,{B_\perp}_i) 
p(\tilde{I}_{ij},U_{ij}|\phi_i,{B_\perp}_i),
\end{align}
where the terms in the likelihood are given by the following normal distributions:
\begin{align}
p(\tilde{I}_{ij},Q_{ij}&|\phi_i,{B_\perp}_i) = \nonumber \\
& \frac{1}{\sqrt{2\pi}\sigma_n} \exp \left[-\frac{ \left(Q_{ij}-
\beta_w {B_\perp}_i^2 \tilde{I}_{ij} \cos \left[ 2 (\phi_i + \phi_{\mathrm{RHT},i}) \right] \right)^2}{2\sigma_n^2} \right], \\
p(\tilde{I}_{ij},U_{ij}&|\phi_i,{B_\perp}_i) = \nonumber \\
& \frac{1}{\sqrt{2\pi}\sigma_n} \exp \left[-\frac{ \left(U_{ij}-
\beta_w {B_\perp}_i^2 \tilde{I}_{ij} \sin \left[ 2 (\phi_i + \phi_{\mathrm{RHT},i}) \right] \right)^2}{2\sigma_n^2} \right].
\end{align}
We note that the product over wavelengths can be computed analytically, and the resulting likelihoods are still normal:
\begin{align}
\mathcal{L}_{Q,i} &= \prod_{j=1}^{N_\lambda} p(\tilde{I}_{ij},Q_{ij}|\phi_i,{B_\perp}_i) \propto \nonumber \\
& \exp \left[ -\frac{1}{2 \sigma_i^2} \left( \frac{S_{QI,i}}{S_{I,i}} - \beta_w {B_\perp}_i^2 \cos \left[ 2 (\phi_i + \phi_{\mathrm{RHT},i}) \right]\right)^2\right] \\
\mathcal{L}_{U,i} &= \prod_{j=1}^{N_\lambda} p(\tilde{I}_{ij},U_{ij}|\phi_i,{B_\perp}_i) \propto \nonumber \\
& \exp \left[ -\frac{1}{2 \sigma_i^2} \left( \frac{S_{UI,i}}{S_{I,i}} - \beta_w {B_\perp}_i^2 \sin \left[ 2 (\phi_i + \phi_{\mathrm{RHT},i}) \right]\right)^2\right],
\end{align}
where
\begin{align}
S_{I,i} = \sum_{j=1}^{N_\lambda} \tilde{I}_{ij}^2, \,\,
S_{QI,i} = \sum_{j=1}^{N_\lambda} Q_{ij} \tilde{I}_{ij}, \,\, S_{UI,i} = \sum_{j=1}^{N_\lambda} U_{ij} \tilde{I}_{ij}, \,\, \sigma_i = \frac{\sigma_n}{\sqrt{S_{I,i}}}
\end{align}

\begin{acknowledgements}
Financial support by the Spanish Ministry of Economy and Competitiveness 
through projects AYA2014-60476-P Consolider-Ingenio 2010 CSD2009-00038
are gratefully acknowledged. AAR also acknowledges financial support
through the Ram\'on y Cajal fellowships. JdlCR is supported by grants from the Swedish Research Council (2015-03994) 
and the Swedish National Space Board (128/15). The Swedish 1-m Solar Telescope is operated on the island of La Palma by the Institute for Solar Physics of Stockholm University in the Spanish Observatorio del Roque de los Muchachos of the Instituto de Astrof\'isica de Canarias.
This research has made use of NASA's Astrophysics Data System Bibliographic Services.
We acknowledge the community effort devoted to the development of the following open-source packages that were
used in this work: \texttt{numpy} (\texttt{numpy.org}), \texttt{matplotlib} (\texttt{matplotlib.org}), \texttt{seaborn} 
(\texttt{stanford.edu/\~\ mwaskom/software/seaborn}),
\texttt{daft} (\texttt{daft-pgm.org}) and \texttt{Stan} (\texttt{https://mc-stan.org}).
\end{acknowledgements}


\end{document}